\documentclass[floatfix,twocolumn,superscriptaddress,pre,nopacs,amsmath,amssymb]{revtex4-2}

\usepackage{amssymb}
\usepackage{amsmath}

\usepackage{color}
\usepackage[pdftex]{graphicx}

\usepackage{empheq}

\usepackage{mathtools}

\usepackage{makecell}

\usepackage{xr}
\makeatletter
\newcommand*{\addFileDependency}[1]{% argument=file name and extension
	\typeout{(#1)}% latexmk will find this if $recorder=0 (however, in that case, it will ignore #1 if it is a .aux or .pdf file etc and it exists! if it doesn't exist, it will appear in the list of dependents regardless)
	\@addtofilelist{#1}% if you want it to appear in \listfiles, not really necessary and latexmk doesn't use this
	\IfFileExists{#1}{}{\typeout{No file #1.}}% latexmk will find this message if #1 doesn't exist (yet)
}
\makeatother

\newcommand*{\myexternaldocument}[1]{%
	\externaldocument{#1}%
	\addFileDependency{#1.tex}%
	\addFileDependency{#1.aux}%
}
\myexternaldocument{supp}

\usepackage{enumerate}
\usepackage{mathtools}
\usepackage{stmaryrd}

\usepackage{enumitem}

\usepackage{textcomp}
\usepackage{gensymb}

\newcommand{\XX}[0]{\boldsymbol{X}}
\newcommand{\KK}[0]{\boldsymbol{K}}

\newcommand{\ex}[1]{\mathrm{e}^{#1}}

\newcommand{\dd}[0]{\mathrm{d}}
\newcommand{\ii}[0]{\mathrm{i}}
\newcommand{\kk}[0]{\boldsymbol{k}}

\newcommand{\rr}[0]{\boldsymbol{r}}

\newcommand{\xx}[0]{\boldsymbol{x}}
\newcommand{\qq}[0]{\boldsymbol{q}}

\newcommand{\yy}[0]{\boldsymbol{y}}

\newcommand{\kB}[0]{k_{\mathrm{B}}}

\newcommand{\cD}[0]{\mathcal{D}}
\newcommand{\bs}[1]{\boldsymbol{#1}}

\newcommand\trick[1]{}

\definecolor{darkblue}{rgb}{0,0,0.6}
\definecolor{darkred}{rgb}{0.6,0,0}
\usepackage[colorlinks=true,urlcolor=darkblue,citecolor=darkblue,linkcolor=darkred]{hyperref}

\newcommand{\nocontentsline}[3]{}
\newcommand{\tocless}[2]{\bgroup\let\addcontentsline=\nocontentsline#1{#2}\egroup}

\begin{document}

\title{Non-Gaussian density fluctuations in the Dean-Kawasaki equation}

\author{Louison Le Bon}
\affiliation{Sorbonne Université, CNRS, Physicochimie des \'Electrolytes et Nanosyst\`emes Interfaciaux (PHENIX), Paris, France}

\author{Antoine Carof}
\affiliation{Universit\'e de Lorraine, CNRS, Laboratoire de Physique et Chimie Th\'eoriques (LPCT), Nancy, France}

\author{Pierre Illien}
\thanks{Corresponding author: pierre.illien@sorbonne-universite.fr}
\affiliation{Sorbonne Université, CNRS, Physicochimie des \'Electrolytes et Nanosyst\`emes Interfaciaux (PHENIX), Paris, France}

\date{\today}

\begin{abstract}

Computing analytically the $n$-point density correlations in systems of interacting particles is a long-standing problem of statistical physics, with a broad range of applications, from the interpretation of scattering experiments in simple liquids, to the understanding of their collective dynamics. For Brownian particles, i.e. with overdamped Langevin dynamics, the microscopic density obeys a stochastic evolution equation, known as the Dean-Kawasaki equation. In spite of the importance of this equation, its complexity makes it very difficult to analyze the statistics of the microscopic density beyond simple Gaussian approximations.  { In this work, resorting to a path-integral description of the stochastic dynamics and relying on a saddle-point analysis in the limit of high density and weak interactions between the particles, we go beyond the usual linearization of the Dean-Kawasaki equation, and we compute exactly the  three- and four-point density correlation functions.} This  result opens the way to using the Dean-Kawasaki equation beyond the simple Gaussian treatments, and could find applications to understand many fluctuation-related effects in soft and active matter systems.

\end{abstract}

\maketitle

\section{Introduction}

Characterizing $n$-point density correlations in systems of interacting particles is a central problem of statistical physics. For instance, measuring two-point density correlations of liquids is generally the simplest insight into their structure and dynamics. Such observables have motivated a significant amount of theoretical work and the development of numerical methods, rooted in classical, equilibrium statistical mechanics ~\cite{McQuarrie1976,Doi1988,Hansen2005}. In addition, they found their applications in the interpretation of neutron, X-ray or light scattering experiments, which have become central tools to analyze liquids, and more generally soft matter ~\cite{Lovesey1986,Berne2000,Stribeck2007,Lindner2024}. More complex observables have also  attracted attention: three-point (or `triplet') correlations have been studied in order to improve the description of liquids beyond simple two-body approaches, and to get better insight into their structure and dynamics~\cite{Alder1964,Rahman1964,Egelstaff1971,Haymet1981,Barrat1988,Attard1989,Coslovich2013a}.

From an analytical perspective, the explicit calculation of $n$-point density correlations is a notoriously difficult problem, and typically require approximations. For instance, to compute static two-point functions, the well-known Percus-Yevick, hypernetted chain or mean spherical closures have been successful~\cite{Hansen2005,Pihlajamaa2024}. Their dynamical counterpart can be estimated using schemes such as mode-coupling theory (MCT)~\cite{Gotze2009}, which remain valid (up to a certain extent) even for very slow dynamics, { and in which high-order density correlations were characterized \cite{VanZon2001}}. However, in spite of their predictive power, such theories almost always rely on numerical evaluations, and fully analytical results are scarce, independently of the underlying dynamics of the system (Newtonian or Langevin).

We will consider the ubiquitous model of Brownian particles which interact via pair potentials -- this level of description is particularly relevant to describe colloidal particles, macromolecular or polymeric fluids, and ions.
The positions of the particles, $\rr_1(t),\dots,\rr_N(t)$ obey overdamped Langevin equations, while the empirical or microscopic density, defined as $\rho(\xx,t) \equiv \sum_{i=1}^N \delta(\xx-\rr_i(t))$, obeys the Dean-Kawasaki (DK) equation~\cite{Kawasaki1994,Dean1996}. The latter is particularly difficult to analyze and has only been studied in specific regimes.
First, in the absence of interactions, the statistics of $\rho$ can be fully characterized as Poissonian (either from direct calculation or from a field-theoretical formulation), as one would expect from simple physical considerations~\cite{Velenich2008}. Second, a one-loop renormalized treatment of the DK equation highlighted its connections with standard MCT theory~\cite{Kim2014} -- this approach is nonetheless not fully consistent, as it is an expansion around a Gaussian theory, and not around the case of noninteracting particles. Finally, another commonly used strategy consists in expanding the DK equation in the limit of high density and weak interactions~\cite{Chavanis2008,Démery2014,Dean2014a}. At leading order, the resulting `linearized' equation, which has been extensively used to study a variety of fluctuation-related phenomena in soft and active matter~\cite{Poncet2021,Poncet2017,Demery2016,Mahdisoltani2021a,Martin2018,Minh2023a}, is by nature limited to describing the Gaussian fluctuations of the stochastic field $\rho$. The characterization of non-Gaussian fluctuations beyond these limiting regimes has been addressed numerically \cite{Bouchet2016}, but remains an open problem from an analytical perspective.

Given the broad range of domains where the DK equation has been studied, from purely theoretical or computational considerations to applications in the physics of soft and active matter for the interpretation of fluctuation-induced phenomena \cite{Illien2024d}, this appears as a prominent line of research, where explicit analytical solutions are still scarce. In this article, resorting to a path-integral formulation and to a perturbative expansion similar to that employed in macroscopic fluctuation theory, we go beyond previous treatments of the DK equation and we compute the {three- and four-point} correlation functions of the microscopic density of interacting Brownian particles. We obtain a simple and fully explicit analytical expression that is confronted to numerical simulations. This constitutes the first analytical characterization of non-Gaussian fluctuations in the DK equation with interactions.

\section{The Dean-Kawasaki equation}
\label{sec_DK}

Consider a $d$-dimensional suspension of $N$ particles interacting via some pair potential $V(\rr)$, and obeying overdamped dynamics. We denote by $\mu$ their bare mobility,  and $D$ their bare diffusion coefficient, which are related through the fluctuation-dissipation relation $D=\mu \kB T$, where $\kB$ is Boltzmann's constant and $T$ the temperature.  Let us start from the set of Langevin equations
\begin{equation}
	\label{overdampedLangevin}
	\frac{\dd \rr_\alpha}{\dd t} = -\mu \sum_{\beta =1}^N \nabla V(\rr_\alpha-\rr_\beta) + \sqrt{2D}\boldsymbol{\zeta}_\alpha(t),
\end{equation}
where $\boldsymbol{\zeta}_\alpha(t)$ are Gaussian white noises, of average zero and variances  $\langle \zeta_{\alpha,i}(t)\zeta_{\beta,j}(t')\rangle =  \delta_{\alpha\beta} \delta_{ij} \delta(t-t')$. The evolution equation of the density of particles, defined as $\rho(\xx,t) \equiv \sum_{\alpha=1}^N \delta(\xx-\rr_\alpha(t))$, reads 
\begin{align}
	&\frac{\partial}{\partial t}\rho(\xx,t) =   \nabla \cdot [\boldsymbol{\xi}(\xx,t) \sqrt{2D\rho(\xx,t)}] +   D\nabla^2 \rho(\xx,t) \nonumber\\
	&+\mu\nabla \cdot \left[ \rho(\xx,t) \int \dd \yy \; \rho(\yy,t)   \nabla V(\xx-\yy) \right]
	\label{DE}
\end{align}
where $\boldsymbol{\xi}(\xx,t)$ is a Gaussian white noise of average zero and variance $\langle \xi_i(\xx,t)\xi_j(\xx',t')\rangle =   \delta_{ij} \delta(\xx-\xx') \delta(t-t').$ This exact evolution equation, usually called the Dean-Kawasaki equation, was derived from phenomenological considerations by the latter~\cite{Kawasaki1994}, and later obtained using stochastic calculus by the former~\cite{Dean1996}. This equation has progressively become an important object of study, as it encompasses under a compact form the $N$-body dynamics described by the coupled overdamped Langevin equations given in Eq.~\eqref{overdampedLangevin}. Its complexity lies in its nonlinearity and in the multiplicative noise term, which makes its direct resolution impossible as such. 

A simple and straightforward treatment of the DK equation consists in linearizing the stochastic density $\rho$ around a constant uniform state $\rho_0=N/\mathcal{V}$ (where $\mathcal{V}=L^d$ is the volume of the system), i.e. writing $ \rho(\xx,t) = \rho_0 + \sqrt{\rho_0}\phi(\xx,t)$, and considering {the joint limit limit $\phi\ll\sqrt{\rho_0}$ and $V\to 0$ with constant $\rho_0 V$ (i.e. the limit of weak interactions and high density)}~\cite{Chavanis2008,Dean2014a,Démery2014}. At leading order, this typically leads to a linear equation obeyed by the perturbation $\phi$ (Appendix \ref{supp_linearization}), that can be solved in Fourier space 
\footnote{Throughout the paper, the convention for Fourier transformation is as follows: 
\begin{equation*}
	\tilde f(\kk,\omega) = \int \dd \rr \;  \ex{-\ii \kk\cdot \rr} \int \dd t \;  \ex{-\ii \omega t} f(\rr,t),
\end{equation*}
and its inverse: 
\begin{equation*}
	 f(\rr,t) = \frac{1}{(2\pi)^d} \int \dd \kk \;  \ex{\ii \kk\cdot \rr} \int \dd \omega \;  \ex{\ii \omega t} f(\kk,\omega).
\end{equation*}\protect\trick.
}
to yield $\tilde\phi(\kk,\omega) = \sqrt{2D}\tilde{\eta}(\kk,\omega) /[\ii \omega+\Omega(\kk)]$, where $\Omega(\kk)$ is homogeneous to an inverse time and is defined as $\Omega(\kk) = D k^2 + \mu\rho_0 k^2 \tilde V(\kk)$, and where  $\tilde{\eta}$ is a Gaussian white noise of zero average and variance $\langle \tilde \eta (\kk,t) \tilde \eta(\kk',t') \rangle= (2\pi)^d k^2 \delta(\kk+\kk')\delta(t-t')$. Therefore, the perturbation $\phi$ is clearly a Gaussian field, and this approximation does not allow the characterization of the non-Gaussian behavior of the field $\rho$. More precisely, within this linearization, { the $n$-point connected correlation functions of $\rho$, i.e. the cumulants of the density,} trivially vanish for $n\geq 3$.

\section{Path-integral formulation}

To go beyond the Gaussian approximation, we follow the path integral formulation of the stochastic dynamics that was initially proposed by Martin-Siggia-Rose~\cite{Martin1973} and Janssen~\cite{Janssen1976}. Its present application is similar to the path integral treatment of the equation of fluctuating hydrodynamics in other contexts~\cite{Krapivsky2014,Krapivsky2015a,Mallick2022,Dandekar2023,Dandekar2024}. The Dean-Kawasaki equation [Eq.~\eqref{DE}] can be rewritten under the form $\partial_t \rho = -\nabla \cdot \boldsymbol{J}$, with the stochastic current $\boldsymbol{J} = -D\nabla \rho -\sqrt{2D\rho} \boldsymbol{\xi} - \mu \rho (\rho \ast \nabla V)$, where we define the convolution operator $(f\ast g)(\xx) = \int \dd \yy f(\yy) g(\xx-\yy)$. 
Since $\boldsymbol{\xi}$ is a unit Gaussian white noise, its probability distribution functional reads $	p[\boldsymbol{\xi}] \propto  \exp\left[-\frac{1}{2} \int_0^{T}\dd t' \int \dd \xx' \boldsymbol{\xi}(\xx',t')^2   \right]$, up to a normalization constant, where ${T}$ is the typical observation time of the trajectory. The probability to observe a given trajectory $\rho$ knowing the initial configuration $\rho(\xx,0)$ then reads
\begin{align}
&	\mathcal{P}[\rho|\rho(\xx,0)] \nonumber\\
& \sim \int \cD \bs{\xi} \; \delta(\partial_t \rho + \nabla\cdot \bs{J})\ex{\left[-\frac{1}{2} \int_0^{T}\dd t' \int \dd \xx' \boldsymbol{\xi}(\xx',t')^2   \right]}
	\label{P1}
\end{align}
up to a normalization prefactor. Note that the DK equation~\eqref{DE} is a stochastic differential equation with multiplicative noise, that is interpreted here in the It\^o way~\cite{Gardiner1985,vanKampen1981}. A consequence is that the Jacobian of the transformation to path-integral is constant, and is absorbed in the normalization~\cite{Zinn-Justin2002,Andreanov2006}. 

Next, we use the Fourier representation of the $\delta$-distribution, which reads, for any functional $\psi[\rho]$: $\delta(\psi[\rho]) = \int \cD \hat\rho \; \exp \left\{ - \int_0^{T} \dd t\int  \dd \xx \hat\rho(\xx,t) \psi[\rho(\xx,t)]  \right\}$, where $\hat\rho$ is an auxiliary field. 
{ With this representation, it is now possible to rewrite the integral as $\mathcal{P}[\rho|\rho(\xx, 0)] \sim \int\mathcal{D}\hat{\rho}\mathcal{D}\boldsymbol{\xi} \exp\left\{-\int_0^T \dd t \, \dd\xx \left[\hat{\rho}(\partial_t\rho + \nabla\cdot\boldsymbol{J}) + \frac{\boldsymbol{\xi}^2}{2}\right]\right\}$. Recalling the definition of $\boldsymbol{J}$, one gets 
\begin{align}
    &\mathcal{P}[\rho|\rho(\xx, 0)]\nonumber\\
    &\sim \int\mathcal{D}\hat{\rho}\ \ex{-\int_0^T dtd\xx\ \hat{\rho}\left(\partial_t\rho - D\nabla^2\rho - \mu\nabla[\rho(\rho*\nabla V)]\right)}G[\rho,\hat{\rho}]
\end{align}
where $G[\hat\rho,{\rho}]= \int\mathcal{D}\boldsymbol{\xi}\ \ex{-\int_0^T\dd t\,\dd\xx\ \left[\frac{\boldsymbol{\xi}^2}{2} +\sqrt{2D\rho}\  \nabla\hat{\rho}\cdot \boldsymbol{\xi}\right]}$, i.e. $G[\hat\rho,{\rho}]=\ex{\int_0^T\dd t\, \dd \xx\ D\rho (\nabla\hat\rho)^2}$.
Finally, performing integration by parts with respect to $\xx$ yields $\mathcal{P}[\rho|\rho(\xx,0)]  \sim \int \cD \hat\rho \; \exp\left\{ - \int_0^{T}\dd t \int \dd \xx\;  S[\rho,\hat\rho]\right\}$
}
with the action
\begin{eqnarray}
	S[\rho,\hat\rho] &\equiv&  \hat \rho \partial_t \rho - D \rho (\nabla \hat \rho)^2 \nonumber\\
&&	+D(\nabla \hat \rho)\cdot(\nabla\rho) +\mu \rho (\nabla \hat \rho) \cdot (\rho\ast\nabla V).
\end{eqnarray}
Incorporating the weight of the initial condition yields $P[\rho]  \sim \int \cD \hat\rho \; \exp\left\{ - \mathcal{S}[\rho,\hat\rho]\right\}$, with the generalized action $\mathcal{S}[\rho,\hat\rho] =- \ln\{P_0[\rho(\xx,t=0)]\} +\int_0^{T}\dd t \int \dd \xx\;  S[\rho,\hat\rho]$, where $P_0$ is the initial distribution of the density.

\section{$n$-point correlation functions}

We now aim at calculating the $n$-point density correlation functions defined in real space as $	C_n(\XX_1,\dots,\XX_n) = \left \langle \prod_{k=1}^n \rho(\XX_k)  \right\rangle_\text{c}$, 
where the index `$\text{c}$' indicates a connected correlation function, and where we introduce the shorthand notation $\XX\equiv(\xx,t)$. They can be computed through the successive functional derivatives of the cumulant generating functional $	\mu[\lambda] =\ln \left \langle \exp\left[ \int \dd \XX \; \lambda(\XX)\rho(\XX)     \right]\right\rangle$ (where the brackets denote average with respect to the distribution $P$), with respect to $\lambda(\XX_1),\dots,\lambda(\XX_n)$, and where we ultimately set $\lambda \equiv 0$.
Equivalently, for $n\geq 2$, the connected correlation functions can be computed as 
\begin{equation}
	\label{Cn_rholamb}
	C_n(\XX_1,\dots,\XX_n) =  \left.\frac{\delta^{n-1}  \langle \rho(\XX_1)\rangle_\lambda }{\delta \lambda(\XX_2) \dots\delta \lambda(\XX_n)}\right|_{\lambda\equiv 0},
\end{equation}
where the average $\langle \cdot\rangle_\lambda$ is taken with respect to the tilted action $	\mathcal{S}_\lambda[\rho,\hat\rho] = -\int \dd \XX \; \lambda(\XX)\rho(\XX) + \mathcal{S}[\rho,\hat\rho]$, i.e. $\langle \cdot\rangle_\lambda = \frac{ \int \cD \rho \int \cD \hat\rho \; \cdot \; \exp( - \mathcal{S}_\lambda[\rho,\hat\rho]) }{ \int \cD \rho \int \cD \hat\rho \; \exp( - \mathcal{S}_\lambda[\rho,\hat\rho])}$.

It is generally not possible to obtain explicit expressions when computing averages with respect to the action $\mathcal{S}_\lambda$, as the integrals cannot be performed easily. { However, one can make progress in the joint limit of high density ($\rho_0\to\infty$) and weak interactions ($V\to 0$), with the product of $\rho_0 V$ being of order $1$ -- this is the limit that was discussed in Section \ref{sec_DK} and that leads to the `linearized' Dean-Kawasaki equation. In this limit, the tilted action can be rewritten as $\mathcal{S}_\lambda [\rho, \hat\rho] = \rho_0 \bar{\mathcal{S}}_\lambda [\bar \rho, \hat\rho]$, where $\bar{\mathcal{S}}_\lambda [\bar \rho, \hat\rho]$ is of order $1$ and independent of $\rho_0$. Applying a saddle-point method, it becomes clear that the average with the tilted action $\mathcal{S}_\lambda$ is dominated by the most probable path, i.e. the path $(\rho,\hat\rho)$ which minimizes the action. This approach is reminiscent of the strategy that is at the heart of macroscopic fluctuation theory (MFT)~\cite{Bertini2001,Bertini2002,Bertini2015}, where saddle point approximations are typically justified by hydrodynamic limits (i.e. long times and large distances).}

We denote the path that minimizes the action by $(q,p)$, and study small variations around it, i.e. we set $\rho(\XX)  =  q(\XX)+ \delta \rho(\XX)$ and $\hat\rho(\XX)  =  p(\XX) + \delta \hat\rho(\XX) $. Defining $\delta \mathcal{S}_\lambda  \equiv \mathcal{S}_\lambda(\rho,\hat\rho)-\mathcal{S}_\lambda(q,p)$ and performing variational calculus, we find that the path of least action, for which $\delta \mathcal{S}_\lambda$ cancels obeys the following equations (Appendix \ref{supp_least_action}):
\begin{eqnarray}
	\partial_t q &=& D\nabla^2 q-2 D\nabla\cdot(q\nabla p) +\mu\nabla \cdot [q (q\ast \nabla V)] \label{bulklamb1} \\
	\partial_t p &=&- D(\nabla p)^2-D\nabla^2 p \nonumber\\
&&	+\mu(\nabla p)\cdot (q\ast \nabla V) -\mu (q\nabla p)\ast\nabla V -\lambda\label{bulklamb2}
\end{eqnarray}
where the Lagrange multiplier $\lambda$ acts as source in the equation for $p$. { These `bulk' equations are completed by initial and final conditions on the values of the fields $q$ and $p$, that are given in Appendix \ref{supp_least_action}}. Although Eqs.~\eqref{bulklamb1}-\eqref{bulklamb2} resemble the MFT equations that were studied for one-dimensional diffusive systems~\cite{Derrida2009,Krapivsky2012,Krapivsky2014,Krapivsky2015a,Mallick2022,Grabsch2022b,Grabsch2023,Dandekar2023,Dandekar2024,Berlioz2024}, they are valid here in $d$ spatial dimensions, and the pair interactions are encoded explicitly in pair potential $V$, rather than in the macroscopic transport coefficients (diffusivity and conductivity) that appear in MFT.

To summarize, the connected density correlation functions are computed using Eq.~\eqref{Cn_rholamb}, i.e. through the successive functional derivatives of $ \langle \rho(\XX_1)\rangle_\lambda$, which is approximated as $q(\XX_1)$ { under the saddle-point approximation}. The function $q$ is  obtained as the solution of the set of equations~\eqref{bulklamb1}-\eqref{bulklamb2}. Importantly, since the determination of $C_n$ through Eq.~\eqref{Cn_rholamb} involves taking $(n-1)$ functional derivatives with respect to $\lambda$ and then taking $\lambda\equiv 0$, one only needs to solve for $q$ at order $\lambda^{n-1}$. To this end, we will introduce the following series expansion $f(\XX) = \sum_{n=0}^\infty f_n(\XX)$ (for $f=p$ or $q$) { where $f_n$ is of order $n$ in $\lambda$}, and solve Eqs.~\eqref{bulklamb1}-\eqref{bulklamb2} order by order.

{
Before solving  Eqs.~\eqref{bulklamb1}-\eqref{bulklamb2}, one needs to specify additional hypotheses and conditions: (i) First, at order $0$ in $\lambda$, the solution of equations Eqs.~\eqref{bulklamb1}-\eqref{bulklamb2} must follow the average evolution of the system. This implies $p_0(\xx,t)=0$ for all $\xx,t$, and we choose to expand the dynamics around the profile $q_0(\xx,t)=\rho_0$, in order to be consistent with the usual expansion of the Dean-Kawasaki around a constant, uniform state, as discussed in Section \ref{sec_DK};
(ii) Second,  we choose to solve these equations in the limit where the system is equilibrated, i.e. when the initial conditions can be sent to the infinite past \cite{Velenich2008}. More rigorously, when computing the $n$-th correlation functions, it means that we take the limit $t_1,\dots,t_n \to \infty $, with finite differences between any two times $t_i-t_j$. In practice, we use Fourier transformations for both space and time to solve Eqs.~\eqref{bulklamb1}-\eqref{bulklamb2}. We emphasize that, in principle, Eqs.~\eqref{bulklamb1}-\eqref{bulklamb2} must be solved for a given initial condition. One can either consider quenched initial conditions (i.e. with a fixed, deterministic initial density profile $ \rho_\text{q}$, implying $P_0[\rho] = \delta(\rho - \rho_\text{q} )$), or annealed (i.e. with a density profile drawn from its equilibrium distribution, implying $P_0[\rho ] = \ex{- \beta \mathcal{F}[\rho]}$, where $\mathcal{F}$ is the free energy of the system) ~\cite{Derrida2009,Krapivsky2014, Krapivsky2015a,Poncet2021a}. The consequences of these choices on the initial and final values of the fields $q$ and $p$ are recalled in Eq. \eqref{ICs}.
(iii) Finally, we emphasize that this approach will yield exact results the limit where $\rho_0$ is very large and where the typical interaction potential $V$ is very small, with the product $\rho_0 V$ being finite.
}

{
As a side remark, we show in Appendix \ref{app_noninteracting} that, starting from Eqs.~\eqref{bulklamb1}-\eqref{bulklamb2}, taking $V=0$, and using a Cole-Hopf transformation \cite{Krapivsky2015a}, one retrieves the exact results for the $n$-point correlation functions of the non-interacting Brownian gas \cite{Velenich2008}.

}

\begin{figure}
	\begin{center}
        \includegraphics[width=0.48\columnwidth]{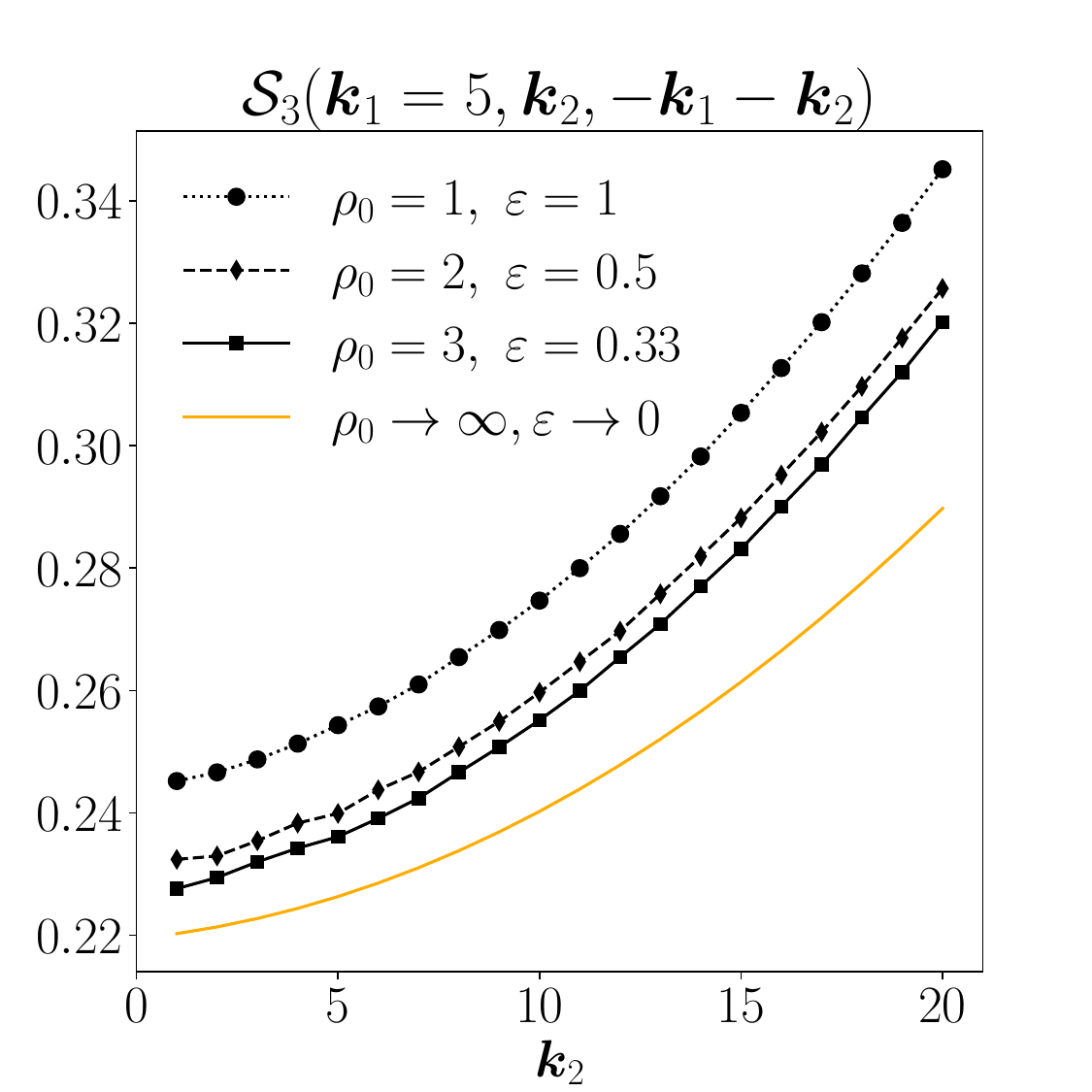}
        \includegraphics[width=0.48\columnwidth]{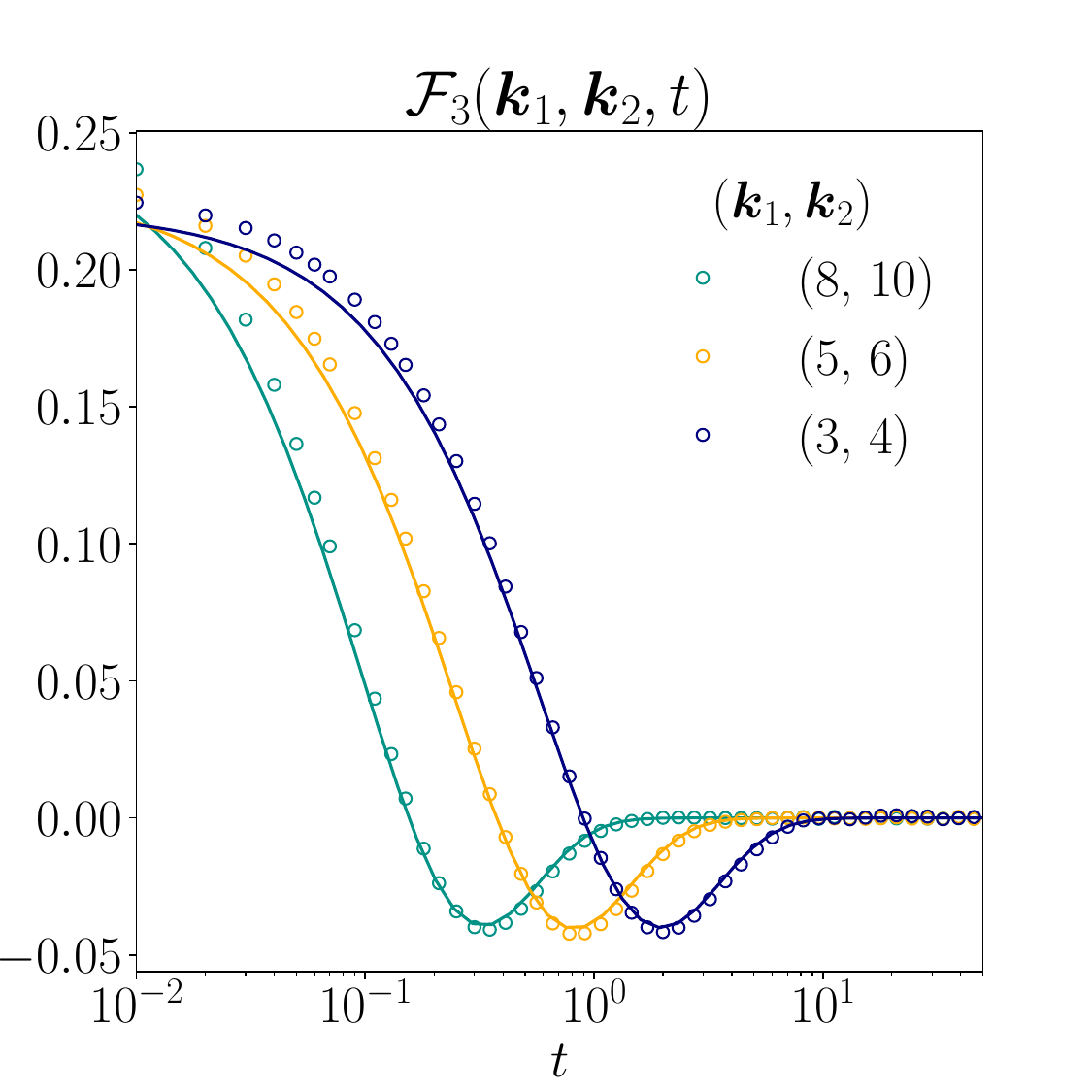}
		\caption{{Results from numerical simulations (symbols) performed in $d=1$, with $N=100$ particles interacting via harmonic repulsion: $V(x) = \varepsilon (1-x/\sigma)^2\Theta(\sigma-x)$ ($\sigma=1$, $\varepsilon=0.5$), compared with the analytical results (solid lines), Eqs.~\eqref{S3} and~\eqref{three_point_two_times} for the left and right panel, respectively. See Appendix \ref{supp_numerical} for details on numerical simulations.  Left: Three-point static structure factor $\mathcal{S}_3(k_1,k_2,-k_1-k_2)$ for $k_1=5$ and as a function of $k_2$. Right: Three-point correlation function $\mathcal{F}_3(k_1,k_2,t) = \frac{1}{N} \langle \tilde{\rho}(k_1,t)\tilde{\rho}(k_2,t)\tilde{\rho}(-k_1-k_2,0) \rangle$ as a function of time, for $(k_1,k_2)=(5,6)$, and for different values of the overall density $\rho_0$. 
        Wavevectors are expressed in units of $2\pi/L$.}}
		\label{fig_F3}
	\end{center}
\end{figure}

\section{Explicit results}

It will be convenient to compute the connected correlation functions in Fourier space, through the relation:
\begin{align}
&\tilde	C_n(\KK_1,\dots,\KK_n) = \left \langle \prod_{i=1}^n \tilde \rho(\KK_i)  \right\rangle_\text{c}\nonumber\\
&	=(2\pi)^{(n-1)(d+1)} \left.	\frac{\delta^{n-1} \tilde q(\KK_1)}{\delta \tilde \lambda(-\KK_2) \dots\delta \tilde \lambda(-\KK_n)}\right|_{\tilde \lambda = 0}
\label{CnFourier}
\end{align}
where we use the shorthand notation $\KK = (\kk,\omega)$. { It will also be convenient to introduce the introduce the quantity $\psi_n$, which is such that:
\begin{align}
 & \tilde{C}_n(\KK_1,\dots,\KK_n)  \nonumber\\
 &= \rho_0(2\pi)^{d+1}\psi_n(\KK_1,\dots, \KK_n)\delta\left(\sum_{i} \KK_i\right).
 \label{def_psi}
\end{align}
}

\subsection{Two-point correlations}

Expanding both $p$ and $q$ in powers of $\lambda$, and at order $1$ in the perturbation, it is straightforward to show that Eqs. ~\eqref{bulklamb1}-\eqref{bulklamb2} yield $\tilde q_1(\KK) = 2D\rho k^2 \mathcal{G}(\KK) \tilde p_1(\KK)$ and $\tilde p_1(\KK) = \mathcal{G}(\KK)^* \tilde \lambda(\KK)$, where we introduce the Green's function $\mathcal{G}(\KK) = [\ii\omega + \Omega(\kk)]^{-1} = [\ii\omega + D k^2 + \mu\rho_0 k^2 \tilde V(\kk)]^{-1}$. Using Eq.~\eqref{CnFourier}, one obtains
\begin{equation}
	\langle  \tilde\rho(\KK_1) \tilde\rho(\KK_2)   \rangle_\text{c}= \frac{2D \rho_0 k_1^2(2\pi)^{d+1}}{\Omega(\kk_1)^2+\omega_1^2} \delta(\kk_1+\kk_2)\delta(\omega_1+\omega_2).
	\label{twopointconnected}
\end{equation}
One check that these are the same correlations as the ones that may be computed within the linearization of the Dean-Kawasaki equation around the homogeneous density $\rho_0$, as described above. Inverting the Fourier transforms with respect to time, Eq.~\eqref{twopointconnected} yields the following expression for the intermediate scattering function, defined for a finite-size system \footnote{Analytical calculations are performed in the thermodynamic limit where $N$, $\mathcal{V}\to\infty$ with a fixed density $\rho_0=N/\mathcal{V}$. Results for finite-size systems, which are required to make comparisons with numerical simulations, can be obtained by making the change $\delta(\kk)\to(\mathcal{V}/(2\pi)^d) \delta_{\kk,0}$.} as $\mathcal{F}_2(k,t) = \frac{1}{N} \langle \tilde \rho(\kk,t) \tilde\rho(-\kk,0)  \rangle = \frac{D k^2}{\Omega(\kk)} \ex{-Dk^2[1+\tilde v(k)]t}$, where we introduce the dimensionless potential $\tilde v (k) \equiv \rho_0 \tilde V(k)/\kB T$. As a consequence, the two-point structure factor that is obtained within this perturbative expansion is $\mathcal{S}(k)=\mathcal{F}_2(k,t=0)=[1+\tilde{v}(k)]^{-1}$, which coincides with the structure factor that is typically computed within the random phase approximation~\cite{Louis2000, Louis2000b, Hansen2005, Démery2014}.

\subsection{Three-point correlations}

We now go one step further, and compute the deviation from the Gaussian behavior, that is encoded in higher-order correlation functions. 
To compute three-point correlation functions, we need to solve Eqs.~\eqref{bulklamb1}-\eqref{bulklamb2} at order $2$ in $\lambda$, which read:
\begin{eqnarray}
	\partial_t q_2  &=&  D\nabla^2 q_2+\mu \rho_0\nabla\cdot (\nabla V\ast q_2) - 2D\rho_0 \nabla^2 p_2 + Q_2 \label{q2_real} \nonumber \\
	\partial_t p_2  &=&  -D\nabla^2 p_2-\rho_0\mu \nabla V\ast \nabla p_2 + P_2 \label{p2_real}
\end{eqnarray}
The source terms $Q_2$ and $P_2$ are explicit in terms of $q_1$ and $p_1$, which both have been calculated when we studied the equations at order $1$ in $\lambda$:
\begin{eqnarray}
Q_2 &=&  -2D\nabla\cdot(q_1\nabla p_1) +\mu \nabla\cdot[q_1(\nabla V\ast q_1)] \\
P_2 &=& -D(\nabla p_1)^2 +\mu (\nabla p_1)\cdot (\nabla V \ast q_1)  - \mu (q_1 \nabla p_1 \ast \nabla V) \nonumber
\end{eqnarray}
{
In Fourier space, the function $q_2$, that is needed to compute three-point correlation functions through Eq.~\eqref{CnFourier}, is formally obtained as $\tilde q_2(\KK) =- 2D\rho_0 k^2 |\mathcal{G}(\KK)|^2 \tilde P_2(\kk,\omega) + \mathcal{G}(\KK) \tilde Q_2(\kk,\omega)$. Using Eq.~\eqref{CnFourier} for $n=3$, and computing the functional derivatives of $\tilde q_2$ with respect to $\tilde \lambda$, we get an explicit expression for the three-point correlation function in Fourier space: $\tilde C_3(\KK_1, \KK_2, \KK_3) =  \rho_0 (2\pi)^{d+1} \delta(\KK_1+\KK_2+\KK_3)\psi_3(\KK_1, \KK_2, \KK_3)$, with  
\begin{align}
    &\psi_3(\KK_{1}, \KK_2, \KK_3) = \ \Gamma_3(\KK_1, \KK_2)\psi_2(\KK_3) \nonumber\\
    &+ \Gamma_3(\KK_1, \KK_3)\psi_2(\KK_2) + \Gamma_3(\KK_2, \KK_3)\psi_2(\KK_1)
    \label{eq:compact_three_point}
\end{align}
where \(\psi_2(\KK)\) is a shorthand for \(\psi_2(-\KK, \KK)\), which is defined from the two-point correlation functions [Eq. \eqref{twopointconnected}] as $\psi_2(\KK,\KK') = -2D (\kk\cdot \kk') \mathcal{G}(\KK)\mathcal{G}(\KK')$, and \(\Gamma_3(\KK, \KK')\) reads:
\begin{align}
    &\Gamma_3(\KK, \KK') = \psi_2(\KK, \KK') \nonumber\\
    &+ \mu\ \kk\cdot \kk'\left[\mathcal{G}(\KK)\tilde{V}(\KK')\psi_2(\KK ') + \KK\leftrightarrow \KK'\right].
\end{align}
}

The Fourier transforms with respect to the frequencies $\omega_1$, $\omega_2$ and $\omega_3$ can all be inverted explicitly, yielding an explicit but lengthy expression for $	\langle \tilde \rho(\kk_1,t_1) \tilde \rho(\kk_2,t_2)  \tilde\rho(\kk_3,t_3)   \rangle_c$, for $t_1<t_2<t_3$  (Appendix \ref{supp_three_point}). To get a simpler expression and to limit the number of variables, we consider the particular situation where $t_1=0$ and $t_2=t_3=t>0$. We get:
{ 
\begin{align}
 &	\langle \tilde \rho(\kk_1,0)  \tilde \rho(\kk_2,t)  \tilde\rho(\kk_3,t)   \rangle_\text{c}  \label{three_point_two_times}\\
 &= \mathcal{S}_3(\kk_1,\kk_2,\kk_3)\Bigg\{\ex{-\Omega(k_1)t}+[1+\tilde{v}(k_1)] \nonumber\\
 & \times \frac{\kk_1\cdot [\kk_2\tilde v(k_2) + \kk_3\tilde v(k_3)](\ex{-[\Omega(k_2) + \Omega(k_3)]t} - \ex{-\Omega(k_1)t})}{[\Omega(k_1) - \Omega(k_2) - \Omega(k_3)]} \Bigg\}\nonumber
\end{align}
}
where we defined the static three-point structure factor as:
\begin{equation}
	\mathcal{S}_3(\kk_1,\kk_2,\kk_3) =  \frac{(2\pi)^d \rho_0 \delta(\kk_1+\kk_2+\kk_3)}{[1+\tilde v(k_1)][1+\tilde v(k_2)][1+\tilde v(k_3)]}.
	\label{S3}
\end{equation}

Eq.~\eqref{three_point_two_times}, which is  exact, is the central result of this Letter, and several comments follow:
(i)~In the absence of interactions ($V=0$), one retrieves the expression of the normalized correlation function that can be derived straightforwardly when the positions of the particles $\rr_i(t)$ are independent Wiener processes~\cite{Velenich2008}, and which reads, for a finite-size system: $\mathcal{F}_3(\kk,\kk',t) \equiv \frac{1}{N}  \langle \tilde\rho(\kk,t) \tilde\rho(\kk',t) \tilde\rho(-\kk-\kk',0) \rangle =  \ex{-D (\kk+\kk')^2 t}$;
(ii)~It is clear that the third-cumulant of the density is different from the second one, meaning that the distribution of $\rho$ is generally non-Poissonian (apart from the special case of noninteracting particles \cite{Velenich2008});
{ (iii)~ The typical relaxation time of the dynamics, that reads $\tau(\kk) = \Omega(\kk)^{-1} = [D k^2 (1+\rho_0 \tilde V(\kk)/\kB T)]^{-1}$ in this limit, controls the relaxation of the three-point correlations. As one could expect, it decreases for increasing density and interactions;
}
(iv)~The expression obtained for the static three-point structure factor [Eq.~\eqref{S3}] coincides with the expression that is usually obtained within the `convolution' or Kirkwood approximation ~\cite{Jackson1962,Ichimaru1970}. It consists in writing $\mathcal{S}_3(\kk_1,\kk_2,\kk_3) \simeq \mathcal{S}(\kk_1) \mathcal{S}(\kk_2) \mathcal{S}(\kk_3)$ where $\mathcal{S}(\kk)$ is the two-point structure factor  that can be proven by writing the three-point extension of the Ornstein-Zernike approximation, and by setting the three-point direct correlation to zero ~\cite{Hansen2005,Barrat1988};
(v)~It can be proven from the analytical expression given in Eq.~\eqref{three_point_two_times} that $\mathcal{F}_3(\kk_1,\kk_2,t) $ is nonmonotonous and always has a negative minimum. This is observed on the plots shown on Fig.~\ref{fig_F3}, where the analytical expression is confronted to results from numerical simulations (see Section~V in~\cite{SM} -- note that, when confronted to numerics, the theory is only expected to be exact asymptotically, for high density and weak interactions). This means that, at long enough times, the density distribution is negatively skewed. This is in contrast with the case of non-interacting particles, where the density if always positively skewed. { Finally, and as expected, we observe that the agreement improves as we approach the joint limit \(\rho_0 \to +\infty\) and \(V \to 0\) while keeping the product \(\rho_0 V\) fixed.}

{
\subsection{Four-point correlations}

We finally go one order further, and derive the expression of four-point connected correlation functions.  This requires the resolution of the equations of motion \eqref{bulklamb1}-\eqref{bulklamb2} at order 3 in $\lambda$, which read:
\begin{align}
    \partial_t q_3 &= D\nabla^2 q_3 +\mu\rho_0\nabla\cdot\left[(q_3*\nabla V)\right]- 2D\rho_0\nabla^2p_3 + Q_3, \nonumber\\
    \partial_t p_3 &=-D\nabla^2p_3-\mu\rho_0(\nabla p_3*\nabla V) + P_3. \label{q3p3}
\end{align}
Here, the source terms \(P_3\) and \(Q_3\) depend on the previously determined fields \(p_1, p_2, q_1\), and \(q_2\): 
\begin{eqnarray}
    Q_3 &&= -2D\nabla\cdot(q_1\nabla p_2 + q_2\nabla p_1) \nonumber\\
    &&+ \mu\nabla\cdot\left[q_1(q_2*\nabla V) + q_2(q_1*\nabla V)\right],\\
    P_3 &&= -2D\nabla p_1\cdot\nabla p_2 + \mu \nabla p_1\cdot (q_2 * \nabla V) \nonumber\\
    &&+ \mu \nabla p_2 \cdot (q_1 * \nabla V) - \mu\left[q_1\nabla p_2 + q_2\nabla p_1\right] * \nabla V.
\end{eqnarray}
In Fourier space, the set of equations \eqref{q3p3} can be solved for both \(\tilde p_3\) and \(\tilde{q}_3\):
\begin{equation}
    \begin{aligned}
        &\tilde q_3(\kk) = -2D\rho_0k^2 |G(\kk)|^2\tilde{P}_3(\kk) + G(\kk)\tilde{Q}_3(\kk)\\
        &\tilde p_3(\kk) = -G(\kk)^*\tilde P_3(\kk)
    \end{aligned}
    \label{eq:EOMkom3rdOrderLambda}
\end{equation}
It should be emphasized that this procedure closely parallels the one used to compute the three-point correlation functions. The structure of equation \eqref{q3p3}, which contains linear terms in \(p_3\) and \(q_3\) and source terms dependent on lower-order \(p_i\) and \(q_i\), is general and expected to hold at arbitrary order. In practice, the main difficulty lies in the rapidly increasing complexity of the source terms \(P_3\) and \(Q_3\), which quickly become intractable. 

From Eq. \eqref{eq:EOMkom3rdOrderLambda}, one can compute the functional derivatives of \(\tilde{q}_3\) with respect to \(\tilde{\lambda}\) to obtain an explicit expression for the fourth cumulant \(\tilde{C}_4\), i.e. applying Eq. \eqref{CnFourier} for $n=4$. Its expression is lengthy but given in a rather compact form in Appendix \ref{app_fourpoint}. A Python notebook is made available: it contains the explicit formula for an arbitrary potential \(V\), together with its Fourier inversion with respect to the frequencies $\omega_i$ \cite{dataset}.

Finally, this expression has been validated by computing numerically the four-point scattering function \(\mathcal{F}_4\), defined as:
\begin{align}
    &\mathcal{F}_4(\kk_1, \kk_2, \kk_3, t)\nonumber\\
    &= \frac{1}{N}\langle\tilde{\rho}(\kk_1, t)\tilde{\rho}(\kk_2, t)\tilde{\rho}(\kk_3, 0)\tilde{\rho}(-\kk_1-\kk_2-\kk_3, 0)\rangle \label{def_F4}
\end{align}
and the static structure factor $\mathcal{S}_4(\kk_1, \kk_2, \kk_3)= \mathcal{F}_4(\kk_1, \kk_2, \kk_3, t=0)$, in the one-dimensional setting already considered in the previous Section. The theoretical and numerical agreement is presented in Fig.~\ref{fig:F4_S4_comp}. Again, like in the results shown on Fig.~\ref{fig_F3}, we observe that the agreement improves as we approach the joint limit \(\rho_0 \to \infty\) and \(\varepsilon \to 0\) with a fixed product. The four-point correlations exhibit a non-trivial non-monotonic time-dependence and may be negative, in contrast with the non-interacting case where it is a positive monotonically decreasing function of time.

\begin{figure}
    \centering
    \includegraphics[width=0.48\linewidth]{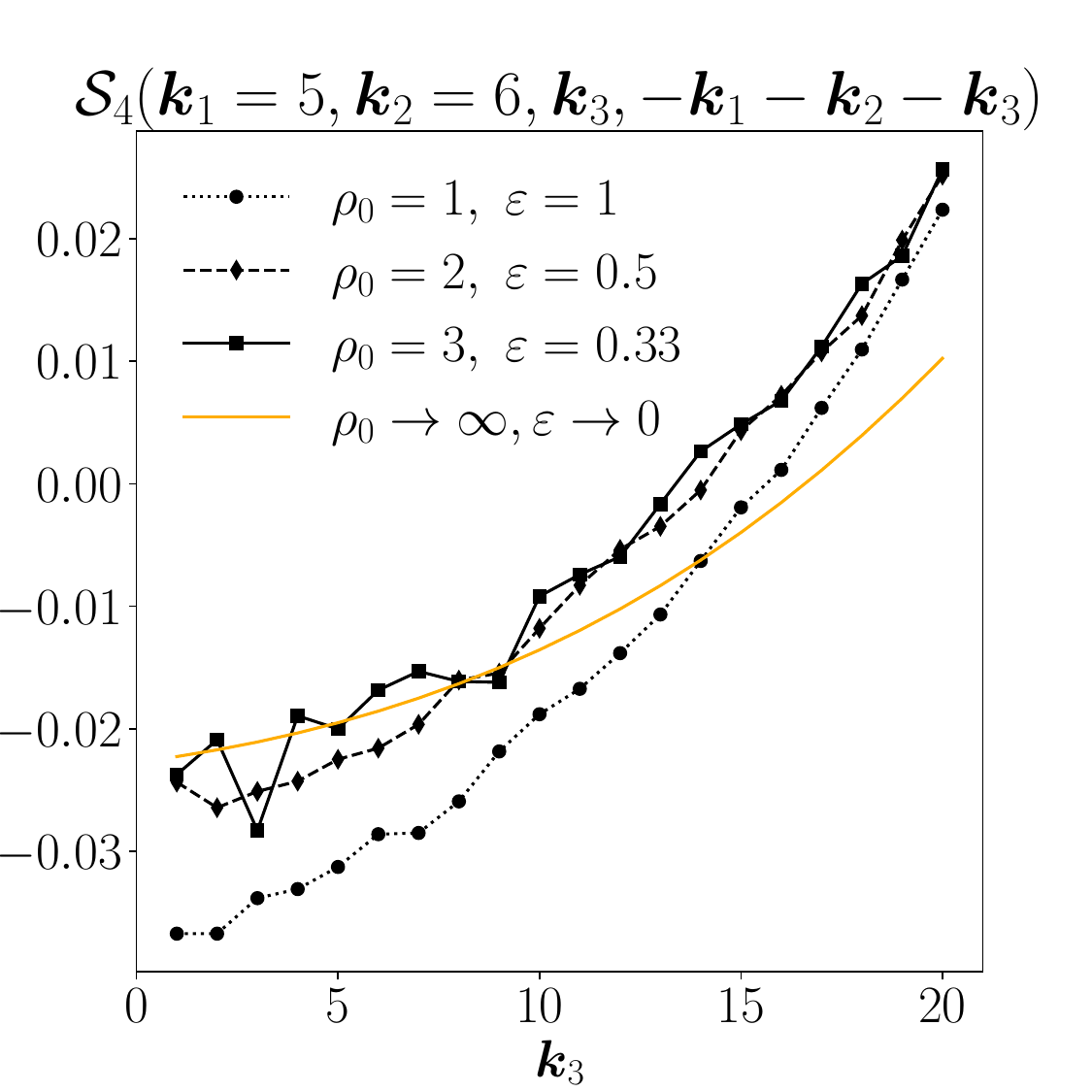}
    \includegraphics[width=0.48\linewidth]{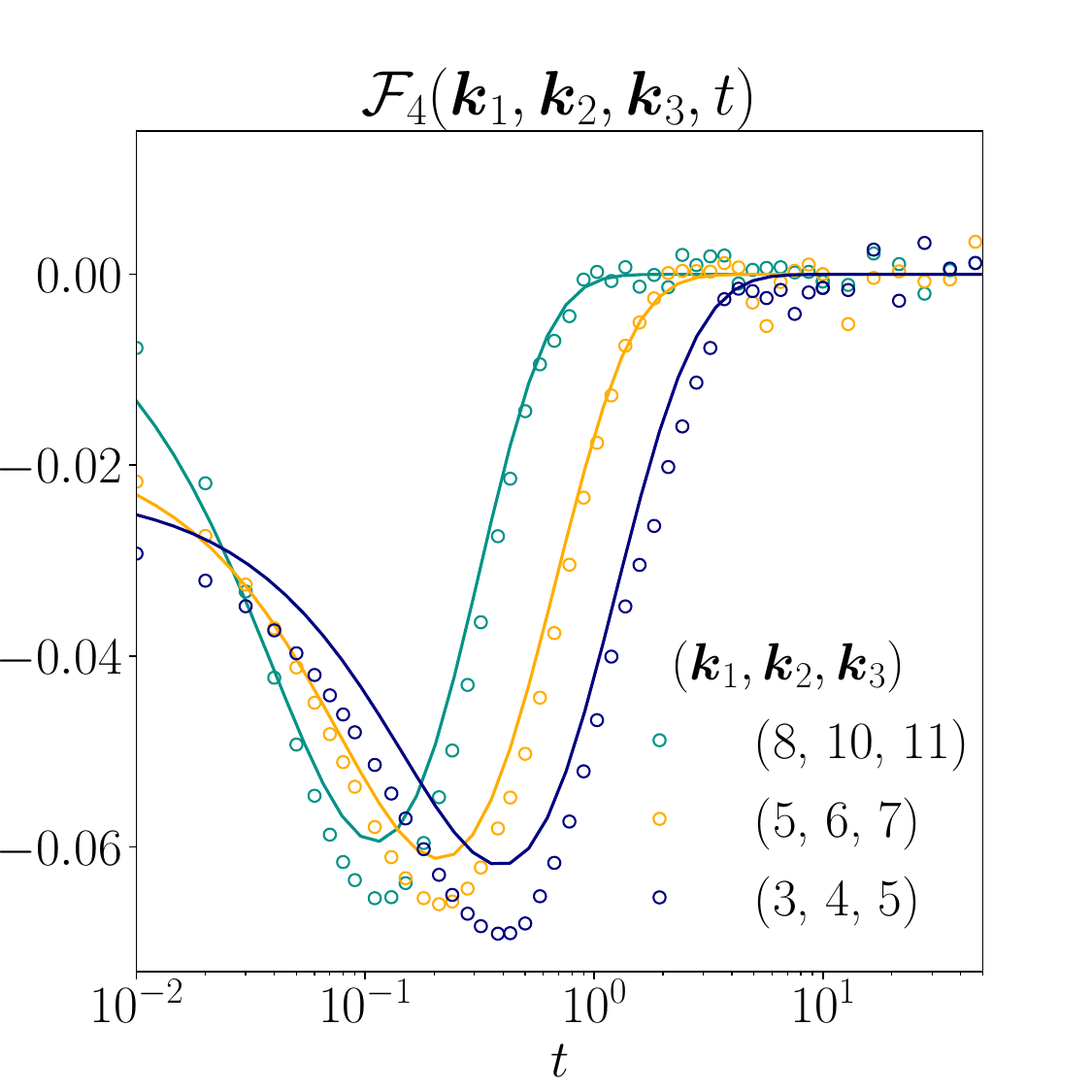}
    \caption{Results from numerical simulations (symbols) performed in $d=1$, compared with the analytical results (see caption of Fig. \ref{fig_F3} and Appendix \ref{supp_numerical} for details on numerical simulations). Left: Four-point static structure factor $\mathcal{S}_4$ for $k_1=5$, $k_2=6$ and as a function of $k_3$. Right: Four-point correlation function $\mathcal{F}_4$ as defined in Eq. \eqref{def_F4} as a function of time, for $(k_1,k_2)=(5,6)$, and for different values of the overall density $\rho_0$. }
    \label{fig:F4_S4_comp}
\end{figure}

}

\section{Conclusion and outlook}

In this work, we provided an analytical description of non-Gaussian density fluctuations in the DK equation. Using a path-integral formulation and macroscopic fluctuation theory, we computed three-point density correlation functions for interacting Brownian particles in the regime of high density and weak interactions. These results, and more generally this methodology, establish the potential of the DK framework to capture higher-order fluctuation phenomena in soft and active matter systems. Looking forward, we aim at computing the full large deviation function of the density, i.e. for arbitrary functional $\lambda$, in the fashion of the full solution of the MFT equations that were obtained in the context of one-dimensional lattice gases~\cite{Grabsch2022b,Mallick2022}. Finally, extensions to  multiple coupled stochastic fields will be crucial to take further the description of electrolytes within the DK framework~\cite{Demery2016,Bonneau2023,Avni2022a,Bernard2023a,Illien2024e}, and more specifically their non-Gaussian fluctuations~\cite{Lesnicki2020,Lesnicki2021}.

\section*{Acknowledgments}

We acknowledge Benjamin Rotenberg, Sophie Hermann, Davide Venturelli, Aurélien Grabsch and Olivier B\'enichou for numerous discussions on this topic. We thank Marie Jardat and Roxanne Berthin for their advice on the computational aspects of the work. P.I. acknowledges financial support from Agence Nationale de la Recherche through project TraNonEq (ANR-24-CE30-0651).

{
\section*{Data availability}

The data and code that support the findings of
this article are openly available \cite{dataset}.

}

\appendix

\onecolumngrid

\section{Reminder: Linearization of the Dean-Kawasaki equation}
\label{supp_linearization}

In this Appendix, we recall the results that are obtained when the Dean-Kawasaki equation [Eq. \eqref{DE} in the main text] is linearized around a homogeneous state. The Dean-Kawasaki equation [Eq. \eqref{DE} in the main text] is nonlinear in the density $\rho$. It can be linearised in the limit of small fluctuations around a constant uniform value $\rho_0$ \cite{Chavanis2008,Dean2014a,Démery2014}: $ \rho(\xx,t) = \rho_0 + \sqrt{\rho_0}\phi(\xx,t)$. Eq. \eqref{DE} then becomes, after having divided both sides by $\sqrt{\rho_0}$): 
\begin{equation}
	\label{LDE}
	\partial_t \phi(\xx,t) = D\nabla^2 \phi(\xx,t) +\mu \rho_0  \nabla [(\phi*\nabla V) (\xx,t)] +\mu \sqrt{\rho_0} \nabla \cdot[\phi (\phi*\nabla V) ] + \sqrt{2D}\nabla \cdot \boldsymbol{\xi}(\xx,t) + \frac{D}{2\sqrt{\rho_0}} \nabla \cdot [ \boldsymbol{\xi}(\xx,t) \phi(\xx,t)],
\end{equation}
with the convolution operator $(V *\phi) (\xx,t) \equiv \int \dd \yy \; V(\xx-\yy)\phi(\yy,t)$. In the limit $\phi\ll\sqrt{\rho_0}$ and $V\to0$ with $\rho_0 V = \mathcal{O}(1)$, two terms may be neglected: the term proportional to $\phi^2$, and the multiplicative noise term. This yields
\begin{equation}
	\label{LDE}
	\partial_t \phi(\xx,t) = D\nabla^2 \phi(\xx,t) + \rho_0 \mu \nabla^2 [(V *\phi) (\xx,t)] + \sqrt{2D}\nabla \cdot \boldsymbol{\xi}(\xx,t).
\end{equation}
In Fourier space, the equation reads
\begin{equation}
	\ii\omega \tilde \phi(\kk,\omega) = -Dk^2\tilde \phi(\kk,\omega) - \mu \rho_0 k^2 \tilde V(k) \tilde \phi(\kk,\omega) + \sqrt{2D} \tilde{{\eta}}(\kk,\omega),
	\label{Fourier_space_phi}
\end{equation}
where the (scalar) noise $\tilde{\eta}$ has zero average and the following variance:
\begin{equation}
	\langle \tilde \eta (\kk,\omega) \tilde \eta(\kk',\omega') \rangle= (2\pi)^{d+1} k^2 \delta(\kk+\kk')\delta(\omega+\omega').
\end{equation}
Solving for $\tilde \phi$, one gets:
\begin{equation}
	\tilde\phi(\kk,\omega) = \frac{\sqrt{2D}}{(\ii \omega+D k^2)+\mu \rho_0 k^2 \tilde V(\kk)}\tilde{{\eta}}(\kk,\omega).
\end{equation}
This is the result given in the main text.\\

We then get the two-point, two-time correlation function:
\begin{equation}
	\langle \tilde\phi(\kk,\omega) \tilde\phi(\kk',\omega') \rangle = \frac{2D}{ \omega^2+[D k^2+\mu \rho_0 k^2 \tilde V(\kk)]^2} (2\pi)^{d+1} k^2 \delta(\kk+\kk')\delta(\omega+\omega')
	%	\label{twoplin}
\end{equation}
The Fourier transform with respect to time can be inverted to yield (for $t>0$):
\begin{equation}
	\label{twoptwotDean}
	\langle \tilde\phi(\kk,t) \tilde\phi(\kk',0) \rangle = (2\pi)^d \delta(\kk+\kk') \frac{D}{ D+\mu \rho_0 \tilde V(\kk)} \ex{-[D k^2+\mu \rho_0 k^2 \tilde V(\kk)]t}
\end{equation}
At equal time, one simply gets
\begin{equation}
	\langle \tilde\phi(\kk,0) \tilde\phi(\kk',0) \rangle
	= (2\pi)^d \delta(\kk+\kk') \frac{1}{ 1+ \rho_0 \tilde V(k)/\kB T},
\end{equation}
which coincides with the expression of the structure factor that can be computed from the random phase approximations (RPA) \cite{Hansen2005}. This is the result that is also obtain from the saddle-point treatment of the path integral formulation of the stochastic dynamics, at order $1$ in $\lambda$, and that is given in the main text.

Finally, given that $\phi$ is Gaussian, all the connected correlation functions or order $n\geq 3$ are zero within this approximation.

\section{Path of least action}
\label{supp_least_action}

In this Appendix, we give details on the derivation of the equations obeyed by the fields $p$ and $q$ that minimize the action $\mathcal{S}_\lambda$, i.e. Eqs. \eqref{bulklamb1} and \eqref{bulklamb2} in the main text. We define $\delta \rho$ and $\delta \hat \rho$, the fluctuations around the optimal path $(q,p)$:
\begin{align}
	\rho(\xx,t) & =  q(\xx,t)+ \delta \rho(\xx,t), \\
	\hat\rho(\xx,t) & =  p(\xx,t) + \delta \hat\rho(\xx,t),
\end{align}
with $\delta \rho, \delta \hat \rho \ll q,p$. Defining $\delta \mathcal{S}_\lambda  \equiv \mathcal{S}_\lambda(\rho,\hat\rho)-\mathcal{S}_\lambda(q,p)$, we find, at leading order in $\delta \rho$ and $\delta \hat \rho$:
\begin{eqnarray}
	\label{deltaS}
	\delta \mathcal{S}_\lambda & \simeq& \int \dd \xx \; \delta\rho(\xx,0) \left[- \frac{\delta \ln P_0[q] }{\delta q(\xx,0)} - p(\xx,0)  \right] + \int \dd \xx\;  \delta\rho(\xx,{T}) p(\xx,{T}) \nonumber\\
	&&+ \int_0^{T} \dd t \int \dd \xx \; \delta \rho(\xx,t) \left\{ -\partial_t p- D(\nabla p)^2-D\nabla^2 p +\mu(\nabla p)\cdot(q\ast\nabla V) -\mu (q\nabla p)\ast\nabla V -\lambda \right\}\nonumber\\
	&&+ \int_0^{T} \dd t \int \dd \xx \; \delta \hat \rho(\xx,t) \left\{ \partial_t q -D\nabla^2 q+2 D\nabla\cdot(q\nabla p) -\mu \nabla \cdot (q (q\ast\nabla V)) \right\}.
\end{eqnarray}

By definition of $(q,p)$, the variation $\delta \mathcal{S}$ given in Eq. \eqref{deltaS} should vanish for all perturbations $\delta \rho$ and $\delta\hat\rho$. The first term in the first line of Eq. \eqref{deltaS} yields two possible conditions, depending on whether one chooses the `annealed' or the `quenched' setting. In the `annealed' setting, $\delta\rho(\xx,0)$ is not specified a priori, so the terms between brackets should vanish. On the other hand, in the `quenched' setting, one has $\delta\rho(\xx,0)=0$. In particular, if we consider the initial configuration to be uniform, we get $q(\xx,0)=\rho_0$. The second term in the first line of Eq. \eqref{deltaS} yields the other boundary condition: $p(\xx,{T})=0$. In summary, the boundary conditions read:
\begin{equation}
	\label{ICs}
	\text{(annealed)}
	\begin{dcases}
		p(\xx,0)=   -\frac{\delta \ln P_0[q]}{\delta q(\xx,0)}  \\
		p(\xx,{T})= 0  
	\end{dcases}
	\qquad \qquad \qquad\qquad
	\text{(quenched)}
	\begin{dcases}
		q(\xx,0)=  \rho_0  \\
		p(\xx,{T})= 0  
	\end{dcases}
\end{equation}
The last two lines in Eq. \eqref{deltaS} give the bulk equations:
\begin{eqnarray}
	\partial_t q &=& D\nabla^2 q-2 D\nabla\cdot(q\nabla p) + \mu\nabla \cdot (q (q\ast\nabla V)), \label{bulk1} \\
	\partial_t p &=&- D(\nabla p)^2-D\nabla^2 p +\mu(\nabla p)\cdot (q\ast\nabla V) -\mu (q\nabla p)\ast \nabla V-\lambda, \label{bulk2}
\end{eqnarray}
whose solution with the boundary conditions specified in Eq. \eqref{ICs} give the path of least { action} and the most probable realisation of the stochastic density obeying the Dean-Kawaski equation. These correspond to Eqs. \eqref{bulklamb1} and \eqref{bulklamb2} in the main text.

{

\section{Noninteracting particles}
\label{app_noninteracting}

In this Appendix, we show that all connected correlation functions can be computed for non-interacting particles, thus retrieving the result of \cite{Velenich2008}. The starting point is the equations of least action \eqref{bulklamb1}-\eqref{bulklamb2}, with the potential \(V\) set to zero. 
The equations of least action read:
\begin{align*}
    \partial_t q &= D\nabla^2q - 2D\nabla\cdot[q\nabla p]\\
    \partial_t p &= -D(\nabla p)^2 - D\nabla^2 p -\lambda
\end{align*}
This set of coupled PDEs can be decoupled using the Cole-Hopf transformation \cite{Krapivsky2015a}, \((Q, P) = (qe^{-p} ,e^p)\), such that \(Q\) (respectively \(P\)) satisfies a diffusion (respectively anti-diffusion) equation with a spatially and temporally dependent growth term \(\lambda\):
\begin{align*}
    \partial_t Q &= D\nabla^2Q  + \lambda Q\\
    \partial_t P &= -D\nabla^2 P - \lambda P
\end{align*}
These equations can be solved order by order in \(\lambda\) by going in \(\kk\)-space, for \(Q_n\) and \(P_n\) which are of order \(n\) in \(\lambda\). Using the boundary conditions \(Q_0(\xx, t) = \rho_0\), \(P_0(\xx, t) = 1\) which correspond to the solutions of the noiseless system, one finds that for \(n > 0\):
\begin{align}
    \tilde Q_{n}(\kk, t) &= \int_0^t ds\ e^{-Dk^2(t - s)}\int\frac{d^q\qq}{(2\pi)^d}\tilde{\lambda}(\kk - \qq, s)\tilde{Q}_{n-1}(\qq, s)
    \label{supp_recursion_Q}\\
    \tilde P_{n}(\kk, t) &= - \int_t^T ds\ e^{+Dk^2(t - s)}\int\frac{d^q\qq}{(2\pi)^d}\tilde{\lambda}(\kk - \qq, s)\tilde{P}_{n-1}(\qq, s)
    \label{supp_recursion_P}
\end{align}
and \(\tilde Q_0(\kk, t) = (2\pi)^d\rho_0\delta(\kk)\), \(\tilde P_0(\kk, t) = (2\pi)^d\delta(\kk)\). These expressions for the fields \(\tilde Q\) and \(\tilde P\) allow one to compute the \(n\)-point connected correlation function \(\tilde C_n\) by adapting Eq. \eqref{CnFourier} from the main text : 
\begin{align*}
    \tilde C_n(\kk_1, \dots, \kk_n;t_1,\dots t_n) &= (2\pi)^{(n-1)d} \frac{\delta^{n-1}\tilde q(\kk_n, t_n)}{\delta\tilde\lambda(-\kk_{1}, t_{1})\cdots\delta\tilde\lambda(-\kk_{n-1}, t_{n-1})}\Bigg|_{\tilde\lambda = 0}\\
    &=\delta_1\cdots\delta_{n-1}\left[\sum_{i=1}^n\int\frac{d^d\qq}{(2\pi)^d}\tilde{P}_i(\kk_n - \qq, t_n)\tilde Q_{n-i}(\qq, t_n)\right]
\end{align*}
where \(\delta_i(...) \equiv (2\pi)^d \delta (...)/\delta \tilde{\lambda}(-\kk_i, t_i)\). Without loss of generality, choosing the time ordering \(0 < t_1 < \cdots < t_n\) ensures that all functional derivatives act only on \(\tilde{Q}\), since \(\tilde{P}_i(\kk, t_n)\) [Eq. \eqref{supp_recursion_P}], viewed as a functional of \(\tilde{\lambda}\), is independent of \(\tilde{\lambda}(\kk, t)\) for \(t < t_n\). This crucial simplification implies that only the \((n-1)\)-th order term \(\tilde{Q}_{n-1}\) in the expansion is needed to compute the \(n\)-point connected correlation function, greatly reducing the complexity of the calculation to:
\[
\tilde{C}_n(\kk_1,\dots,\kk_n; t_1,\dots, t_n) = (2\pi)^{(n-1)d}\frac{\delta^{n-1}\tilde{Q}_{n-1}(\kk_n, t_n)}{\delta\tilde\lambda(-\kk_{n-1}, t_{n-1})\cdots\delta\tilde\lambda(-\kk_1, t_1)}.
\]
The situation is now the opposite of that for \(\tilde{P}\). Indeed, for \(\tilde{Q}_{n-2}(\qq, s)\) (See Eq.\eqref{supp_recursion_Q}), the integrals involved span times from \(0\) to \(t_{n-2}\), implying that functional derivatives with respect to \(\tilde{\lambda}(-\kk_i, t_i)\) vanish for any \(t_i > t_{n-2}\). Consequently, the only non-vanishing contributions arise from the combination \(\delta_{n-1}[\tilde\lambda(\kk_n - \qq, s)]\delta_{n-2}\cdots\delta_1[\tilde Q_{n-1}(\qq, s)]\). This structure leads to a recursive formula for the connected \(n\)-point correlation functions of the perfect Brownian gas:
\begin{equation}
    \tilde C_n(\kk_1, \dots, \kk_n;t_1,\dots t_n) = e^{-Dk_n^2(t_n - t_{n-1})}\tilde{C}_{n-1}(\kk_1,\dots, \kk_{n-1} + \kk_{n}; t_1,\dots, t_{n-1}).
    \label{recursion_no_int}
\end{equation}
The two-point correlation function, which sets the initial condition for the recursive formula, reads:
\[
\tilde{C}_2(\kk_1,\kk_2;t_1,t_2) = \rho_0\ (2\pi)^d\delta(\kk_1 + \kk_2)\, \exp\left[D\, \kk_1 \cdot \kk_2\, (t_2 - t_1)\right].
\]
Using this recursive formula, one can derive an explicit expression for the \(n\)-point connected correlation function:
\begin{equation}
    \langle\tilde\rho(\kk_1, t_1)\cdots\tilde\rho(\kk_n, t_n)\rangle_c= (2\pi)^d\delta\left(\sum_i\kk_i\right)\rho_0\exp\left\{D\sum_{i <j} \kk_i\cdot\kk_j (t_j - t_i)\right\}
\end{equation}
Although our derivation was done in the limit $\rho_0 \to \infty$, the final result is valid at arbitrary density \cite{Velenich2008}.

As a concluding remark regarding the noninteracting Brownian gas, we note that the recursion relation \eqref{recursion_no_int} can be Fourier transformed back to real space. This yields the expression of the \(n\)-point connected correlation function for non-interacting particles in real space, for \(0 < t_1 <\dots<t_n\):
\begin{align}
    C_n(\xx_1,\ldots,\xx_n) &= P(x_1, t_1)\prod_{i=1}^nP(\xx_{i+1}, t_{i+1} |\xx_{i}, t_{i})\label{eq:BrownCorr xt}\\
    &= P[\xx(t_1)=\xx_1,\ldots, \xx(t_n) = \xx_n]\notag
\end{align}
where \(P(\xx_1, t_1) = N/V = \rho_0\) and \(P(\xx', t' \mid \xx, t)\) is the propagator of a single particle, interpreted as the probability density of finding the particle at position \(\xx'\) at time \(t'\), given that it was at position \(\xx\) at time \(t\). The last line holds for any Markovian stochastic process. In the case of Brownian motion, and assuming \(t' > t\), the propagator takes the Gaussian form $P(\xx', t' | \xx, t) = \left[4\pi D (t' - t)\right]^{-d/2}\exp\left\{-\frac{(\xx' - \xx)^2}{4D(t' - t)}\right\}$, and yields the real-space expression from Ref. \cite{Velenich2008}.

}

\section{Derivation of the three-point correlation functions}
\label{supp_three_point}

In this Appendix, we give details on the derivation of Eq. \eqref{three_point_two_times} in the main text.

\subsection{Expression in Fourier space}

The starting point of this derivation is the expansion of Eqs. \eqref{bulklamb1}-\eqref{bulklamb2} at order $2$ in $\lambda$. They read:
\begin{eqnarray}
	\partial_t q_2 & = & D\nabla^2 q_2+\mu \rho_0\nabla\cdot (\nabla V\ast q_2) - 2D\rho_0 \nabla^2 p_2\underbrace{ -2D\nabla\cdot(q_1\nabla p_1) +\mu \nabla\cdot[q_1(\nabla V\ast q_1)]}_{\equiv Q_2}, \label{q2_real}\\
	\partial_t p_2 & = & -D\nabla^2 p_2-\rho_0\mu \nabla V\ast \nabla p_2\underbrace{-D(\nabla p_1)^2 +\mu (\nabla p_1)\cdot (\nabla V \ast q_1) - \mu (q_1 \nabla p_1 \ast \nabla V) }_{\equiv P_2},\label{p2_real}
\end{eqnarray}
where $Q_2$ and $P_2$ play the roles of sources. In Fourier space, the first equation yields the following expression of $\tilde q_2$:
\begin{equation}
	\tilde q_2(\kk,\omega) = \frac{2D \rho_0 k^2}{ \ii \omega+\Omega(k) } \tilde p_2(\kk,\omega)+\frac{\tilde Q_2(\kk,\omega)}{\ii\omega +\Omega(\kk)}.
\end{equation}
From Eq. \eqref{p2_real}, one gets $\tilde p_2(\kk,\omega)=\tilde P_2(\kk,\omega)/(\ii\omega - \Omega(\kk))$. Finally, $\tilde q_2$ is given explicitly in terms of $\tilde P_2$ and $Q_2$:
\begin{equation}
	\tilde q_2(\kk,\omega) =- \frac{2D\rho_0 k^2}{\Omega(\kk)^2+\omega^2}\tilde P_2(\kk,\omega) + \frac{\tilde Q_2(\kk,\omega)}{\ii\omega+\Omega(\kk)},
	\label{q2_P2Q2}
\end{equation}
which are themselves explicit functions of the functions $p_1$ and $q_1$ determined explicitly at the previous step of the calculation. The goal of the calculation is to compute the three-point correlation functions:
\begin{eqnarray}
	\tilde C_3(\kk_1,\kk_2,\kk_3,\omega_1,\omega_2,\omega_3) &=&  \langle \tilde \rho(\kk_1,\omega_1)  \rho(\kk_2,\omega_2)  \rho(\kk_3,\omega_3)   \rangle_\text{c},\\
	&=&(2\pi)^{2(d+1)} 
	\left.	\frac{\delta^{2} q(\kk_1,\omega_1)}{\delta \tilde \lambda(-\kk_2,-\omega_2) \delta \tilde \lambda(-\kk_3,-\omega_3)}\right|_{\tilde \lambda = 0}, \\
	&=&(2\pi)^{2(d+1)} 
	\frac{\delta^{2} q_2(\kk_1,\omega_1)}{\delta \tilde \lambda(-\kk_2,-\omega_2) \delta \tilde \lambda(-\kk_3,-\omega_3)} .
\end{eqnarray}
Using Eq. \eqref{q2_P2Q2}, one gets:
\begin{align}
	&\langle \tilde \rho(\kk_1,\omega_1)  \rho(\kk_2,\omega_2)  \rho(\kk_3,\omega_3)   \rangle_\text{c} \nonumber\\
	&= (2\pi)^{2(d+1)}  \left\{  - \frac{2D\rho_0 k_1^2}{\Omega(\kk_1)^2+\omega_1^2}  \frac{\delta^{2} \tilde P_2(\kk_1,\omega_1)}{\delta \tilde \lambda(-\kk_2,-\omega_2) \delta \tilde \lambda(-\kk_3,-\omega_3)} 	+ \frac{1}{\ii\omega_1+\Omega(\kk_1)}  \frac{\delta^{2} \tilde Q_2(\kk_1,\omega_1)}{\delta \tilde \lambda(-\kk_2,-\omega_2) \delta \tilde \lambda(-\kk_3,-\omega_3)}
	\right\},
	\label{C3_deriv}
\end{align}
From the definitions of $Q_2$ and $P_2$ (in Eqs. \eqref{q2_real} and \eqref{p2_real} respectively), one can compute their Fourier transforms as:
\begin{eqnarray}
	\tilde Q_2(\kk,\omega) &=& \int \frac{\dd \kk'\dd\omega'}{(2\pi)^{d+1}} (\kk-\kk')\cdot \kk \; \tilde q_1(\kk',\omega') \left[ 2D \tilde p_1(\kk-\kk',\omega-\omega') - \mu \tilde V(\kk-\kk') \tilde q_1(\kk-\kk',\omega-\omega')\right], \\
	\tilde P_2(\kk,\omega) &=& D \int \frac{\dd \kk' \dd \omega'}{(2\pi)^{d+1}} \kk' \cdot (\kk-\kk') \tilde p_1(\kk',\omega') \tilde p_1(\kk-\kk',\omega-\omega') \nonumber \\
	&& -\mu  \int \frac{\dd \kk' \dd \omega'}{(2\pi)^{d+1}} \kk' \cdot (\kk-\kk')\tilde p_1(\kk',\omega') \tilde  V(\kk-\kk') q_1(\kk-\kk',\omega-\omega') \nonumber\\
	&&+ \mu \int \frac{\dd \kk' \dd \omega'}{(2\pi)^{d+1}} \kk \cdot (\kk-\kk')\tilde q_1(\kk',\omega') \tilde V(\kk) p_1(\kk-\kk',\omega-\omega').
\end{eqnarray}
Recalling for completeness the expressions of $\tilde p_1$ and $\tilde q_1$:
\begin{align}
	\tilde p_1(\kk,\omega)& =\frac{ \tilde \lambda(\kk,\omega)}{\Omega(\kk)-\ii\omega},\\
	\tilde q_1(\kk,\omega) &= \frac{2D\rho k^2 \tilde\lambda(\kk,\omega)}{\Omega(\kk)^2+\omega^2},
\end{align}
one then computes the second functional derivatives of $\tilde Q_2$ and $\tilde P_2$, which read:
\begin{align}
	&\frac{\delta^{2} \tilde Q_2(\kk_1,\omega_1)}{\delta \tilde \lambda(-\kk_2,-\omega_2) \delta \tilde \lambda(-\kk_3,-\omega_3)} = \frac{1}{(2\pi)^{d+1}} \frac{2D\rho_0 k_2^2}{\Omega(k_2)^2+\omega_2^2} \kk_1\cdot (\kk_1+\kk_2) \delta(\kk_1+\kk_2+\kk_3 )\delta(\omega_1+\omega_2+\omega_3) \nonumber\\
	&\times \left[ \frac{2D}{\Omega(\kk_1+\kk_2)-\ii (\omega_1+\omega_2)} - \mu \tilde V(\kk_1+\kk_2) \frac{2D\rho_0(\kk_1+\kk_2)^2}{\Omega(\kk_1+\kk_2)^2+(\omega_1+\omega_2)^2}\right] + 2\leftrightarrow 3 ,
\end{align}
and
\begin{align}
	&\frac{\delta^{2} \tilde P_2(\kk_1,\omega_1)}{\delta \tilde \lambda(-\kk_2,-\omega_2) \delta \tilde \lambda(-\kk_3,-\omega_3)} = \frac{1}{(2\pi)^{d+1}}\delta(\kk_1+\kk_2+\kk_3)\delta(\omega_1+\omega_2+\omega_3) \left\{  \frac{D\kk_2\cdot(\kk_1+\kk_2)}{[\Omega(\kk_2)+\ii\omega_2][\Omega(\kk_1+\kk_2)-\ii(\omega_1+\omega_2)]} \right. \nonumber\\
	&+\mu \kk_2\cdot(\kk_1+\kk_2) \tilde V(\kk_1+\kk_2) \frac{2D\rho_0(\kk_1+\kk_2)^2}{[\Omega(\kk_2)+\ii\omega_2][\Omega(\kk_1+\kk_2)^2+(\omega_1+\omega_2)^2]}\nonumber\\
	&\left. +\mu \kk_1\cdot(\kk_1+\kk_2) \tilde V(\kk_1) \frac{2D\rho_0 \kk_2^2}{[\Omega(\kk_1+\kk_2)-\ii(\omega_1+\omega_2)][\Omega(\kk_2)^2+\omega_2^2]} +2\leftrightarrow 3 \right\}.
\end{align}
Finally, with Eq. \eqref{C3_deriv}, one gets:
\begin{align}
	&	\langle \tilde \rho(\kk_1,\omega_1)  \rho(\kk_2,\omega_2)  \rho(\kk_3,\omega_3)   \rangle_\text{c}= (2\pi)^{d+1} \delta(\kk_1+\kk_2+\kk_3)\delta(\omega_1+\omega_2+\omega_3) \nonumber\\
	&\times \left\{   \frac{2D^2\rho_0(\kk_1+\kk_2)}{\Omega(\kk_1+\kk_2)-\ii(\omega_1+\omega_2)}  \cdot\left( \frac{k_1^2 \kk_2}{[\Omega(\kk_2)+\ii\omega_2][\Omega(\kk_1)^2+\omega_1^2]}
	+\frac{2k_2^2 \kk_1}{[\Omega(\kk_1)+\ii\omega_1][\Omega(\kk_2)^2+\omega_2^2]}  
	\right) + 2\leftrightarrow 3   \right. \nonumber\\
	&-\mu(2D\rho_0)^2 \frac{\tilde V(\kk_1+\kk_2) (\kk_1+\kk_2)^2 (\kk_1+\kk_2)}{\Omega(\kk_1+\kk_2)^2+(\omega_1^2+\omega_2^2)}\cdot\left( \frac{k_1^2 \kk_2}{[\Omega(\kk_2)+\ii\omega_2][\Omega(\kk_1)^2+\omega_1^2]}
	+\frac{k_2^2 \kk_1}{[\Omega(\kk_1)+\ii\omega_1][\Omega(\kk_2)^2+\omega_2^2]}  
	\right) + 2\leftrightarrow 3 \nonumber\\
	&\left.-\mu(2D\rho_0)^2 \frac{\tilde V(\kk_1) k_2^2 k_1^2 \kk_1\cdot(\kk_1+\kk_2)}{[\Omega(\kk_1+\kk_2)-\ii(\omega_1+\omega_2)][\Omega(\kk_1)^2+\omega_1^2][\Omega(\kk_2)^2+\omega_2^2]}   + 2\leftrightarrow 3 \right\}\\
	&\equiv {\rho_0}(2\pi)^{d+1} \delta(\kk_1+\kk_2+\kk_3)\delta(\omega_1+\omega_2+\omega_3) \psi_3(\kk_1,\kk_2,\kk_3,\omega_1,\omega_2,\omega_3). \label{def_psi3}
\end{align}
The last equality defines $\psi_3$. This is the most general expression for the connected three-point correlation function. In order to simplify this expression, we introduce the shorthand notation $\KK_i\equiv (\kk_i, \omega_i)$ and:
\begin{equation}
	\mathcal{G}(\KK_i) = \frac{1}{\Omega(\kk_i) + \ii\omega_i}.
\end{equation}
{
One gets the following expression for \(\psi_3\):
\begin{equation}
    \psi_3(\KK_{1}, \KK_2, \KK_3) = \ \Gamma_3(\KK_1, \KK_2)\psi_2(\KK_3) + \Gamma_3(\KK_1, \KK_3)\psi_2(\KK_2) + \Gamma_3(\KK_2, \KK_3)\psi_2(\KK_1),
    \label{eq:compact_three_point}
\end{equation}
where \(\psi_2(\KK)\) is a shorthand for \(\psi_2(-\KK, \KK)\) and \(\Gamma_3(\KK, \KK')\) reads:
\begin{align}
    \Gamma_3(\KK, \KK') &\equiv -2D\ \kk\cdot \kk'\ G(\KK) G(\KK') + \mu\ \kk\cdot \kk'\left[G(\KK)\tilde{V}(\KK')\psi_2(\KK ') + \KK\leftrightarrow \KK'\right]\\
    &= \psi_2(\KK, \KK') + \mu\ \kk\cdot \kk'\left[G(\KK)\tilde{V}(\KK')\psi_2(\KK ') + \KK\leftrightarrow \KK'\right]
\end{align}
As written, \eqref{eq:compact_three_point}, the expression of the three-point correlation function is explicitly symmetric under the exchange \(\KK_i\leftrightarrow \KK_j\).

}

\subsection{Expression in real time}

We next invert the Fourier transforms with respect to time, in order to compute:
\begin{equation}
	\langle \tilde \rho(\kk_1,t_1)  \rho(\kk_2,t_2)  \rho(\kk_3,t_3)   \rangle_\text{c}
	= \int_{-\infty}^\infty \frac{\dd \omega_1}{2\pi} \ex{\ii\omega_1 t_1}
	\int_{-\infty}^\infty \frac{\dd \omega_2}{2\pi} \ex{\ii\omega_2 t_2}
	\int_{-\infty}^\infty \frac{\dd \omega_3}{2\pi} \ex{\ii\omega_3 t_3}	\langle \tilde \rho(\kk_1,\omega_1)  \rho(\kk_2,\omega_2)  \rho(\kk_3,\omega_3)   \rangle_\text{c}.
\end{equation}
Using the fact the correlation function is in fact proportional to $\delta(\omega_1+\omega_2+\omega_3)$, we get, using the definition of $\psi_3$ [Eq. \eqref{def_psi3}]:
\begin{eqnarray}
	\langle \tilde \rho(\kk_1,t_1)  \rho(\kk_2,t_2)  \rho(\kk_3,t_3)   \rangle_\text{c}
	&=&  (2\pi)^{d} \delta(\kk_1+\kk_2+\kk_3)\nonumber\\
	&&\times	\int_{-\infty}^\infty \frac{\dd \omega_1}{2\pi} \ex{\ii\omega_1 (t_1-t_3)}
	\int_{-\infty}^\infty \frac{\dd \omega_2}{2\pi} \ex{\ii\omega_2 (t_2-t_3)}
	\psi_3(\kk_1,\kk_2,\kk_3,\omega_1,\omega_2,-\omega_1-\omega_2)
\end{eqnarray}
One can get a general expression without making any assumption of the order of the times $t_1$, $t_2$ and $t_3$, i.e. by defining quantities like $\sigma_{ab} = \text{sgn}(t_a-t_b)$, and expressing the final result in terms of $\sigma_{ab}$. { To simplify, we make the choice $0<t_1<t_2<t_3$, and the result reads:
\begin{align}
		&\langle \tilde \rho(\kk_1,t_1) \tilde \rho(\kk_2,t_2)  \tilde\rho(\kk_3,t_3)   \rangle_c ={\rho_0D}\Bigg\{\left[\frac{2Dk_2^2 \kk_3\cdot(\kk_1v_1 + \kk_2 v_2)}{\Omega_1 + \Omega_2 - \Omega_3} - 2\kk_2\cdot \kk_3\right] \frac{e^{-(t_2 - t_1) \Omega_1 - (t_3 - t_2) \Omega_3}}{(\Omega_2 - \Omega_1 + \Omega_3)(1 + v_1)} \nonumber  \\
		&+ \frac{\kk_3 \cdot (\kk_1 v_1 + \kk_2 v_2)}{(1 + v_1)(1 + v_2)} \frac{e^{- (t_3 - t_1) \Omega_1 -(t_3 - t_2) \Omega_2}}{\Omega_3-\Omega_2 - \Omega_1} + \frac{\kk_1 \cdot (\kk_2 v_2 + \kk_3 v_3)}{(1+v_2)(1 + v_3)} \frac{e^{-(t_2 - t_1) \Omega_2 - (t_3 - t_1) \Omega_3}}{\Omega_1 - \Omega_2 - \Omega_3}\nonumber \\
		&+ \frac{\kk_2 \cdot (\kk_1v_1 +\kk_3v_3)}{(1+v_1)(1+v_3)}\frac{e^{-(t_2 - t_1) \Omega_1 - (t_3 - t_2) \Omega_3}}{\Omega_2 - \Omega_1 + \Omega_3}\Bigg\} \nonumber
	\label{eq:3point kt}
\end{align}
where we use the shorthand notations $	v_j = \frac{\rho_0 \tilde V(\kk_j)}{\kB T}$ and $\Omega_j = Dk^2\left(1+\frac{\rho_0 \tilde V(\kk_j)}{\kB T}\right)$.}
Finally, taking $t_2=t_3=t>0$ and $t_1=0$, we get Eq. \eqref{three_point_two_times} in the main text.

{
\section{Expression of the four-point correlation function}
\label{app_fourpoint}

In this Appendix, we give details on the expression of the four-point correlation function \(\tilde{C}_4(\KK_1,\KK_2, \KK_3, \KK_4)\) whose expression is provided in the variables \((\kk, t)\) with times ordered as \(0 < t_1 < t_2 < t_3 < t_4\) in the associated Python notebook \cite{dataset}. To this end, we first recall the expressions of the two- and three-point correlation functions, and then show how they relate to the four-point correlation function.

First, we recall the definition of the amplitude \(\psi_n\) associated with the \(n\)-point correlation function \(\tilde{C}_n\) (see Eq. \eqref{def_psi} in the text):
\begin{equation}
\tilde{C}_n(\KK_1,\dots,\KK_n) = \rho_0(2\pi)^{d+1}\psi_n(\KK_1,\dots, \KK_n)\delta\left(\sum_{i} \KK_i\right),
\end{equation}
such that the amplitude \(\psi_2\) associated with the two-point correlation function \(\tilde{C}_2\), given by Eq.\eqref{twopointconnected} in the main text, reads:
\begin{equation}
\psi_2(\KK_1, \KK_2) = -2D\ \kk_1\cdot\kk_2\ \mathcal{G}(\KK_1) \mathcal{G}(\KK_2)
\end{equation}
With this expression, \(\psi_3\), given by Eq.\eqref{eq:compact_three_point}, can be written as a function of \(\psi_2\):
\begin{equation}
    \psi_3(\KK_{1}, \KK_2, \KK_3) = \ \Gamma_3(\KK_1, \KK_2)\psi_2(\KK_3) + \Gamma_3(\KK_1, \KK_3)\psi_2(\KK_2) + \Gamma_3(\KK_2, \KK_3)\psi_2(\KK_1)
\end{equation}
Here, \(\psi_2(\KK)\) denotes \(\psi_2(-\KK, \KK)\), and the function \(\Gamma_3(\KK, \KK')\) is given by:
\begin{equation}
\Gamma_3(\KK, \KK') = \psi_2(\KK, \KK') + \mu\ \kk\cdot \kk'\left[\mathcal{G}(\KK)\tilde{V}(\kk')\psi_2(\KK ') + \KK\leftrightarrow \KK'\right]
\end{equation}
Finally, following the same steps as in Appendix~\ref{supp_three_point} and expanding up to third order in \(\lambda\), one obtains the four-point correlation function. Although the calculation is cumbersome, the result reads:
\begin{align}
    \psi_4(\KK_1,\KK_2, \KK_3, \KK_4) =&\hspace{0.3cm}\Gamma_4(\KK_1, \KK_4)\psi_3(\KK_1 +\KK_4, \KK_2, \KK_3)\label{eq:compact_four_point}\\
    &\begin{aligned}
        +\Gamma_4(\KK_2, \KK_3) \Big\{&\Gamma_2(\KK_1, \KK_2 + \KK_3)\psi_2(\KK_4) +\Gamma_2(\KK_4, \KK_2 + \KK_3)\psi_2(\KK_1)\\
        &+\mu\ \kk_1\cdot(\kk_2 + \kk_3)\ \tilde{V}(\kk_1)\mathcal{G}(\KK_2 +\KK_3)\psi_2(\KK_1)\psi_2(\KK_4)\\
        &+\mu\ \kk_4\cdot(\kk_2 + \kk_3)\ \tilde{V}(\kk_4)\mathcal{G}(\KK_2 +\KK_3)\psi_2(\KK_1)\psi_2(\KK_4)\Big\}
    \end{aligned}\notag\\
    &+\ 4\leftrightarrow 2 +\ 4\leftrightarrow 3\notag,
\end{align}
where the function \(\Gamma_4\) is defined as:
\begin{equation}
\Gamma_4(\KK, \KK') \equiv \Gamma_3(\KK, \KK') - \mu\ \tilde{V}(\kk + \kk')\ (\kk + \kk')\cdot\left[\kk\ \mathcal{G}(\KK)\psi_2(\KK') + \KK\leftrightarrow\KK'\right].
\end{equation}
As written, the four-point correlation function~\eqref{eq:compact_four_point} is not explicitly symmetric under the exchange \(\KK_i\leftrightarrow\KK_j\), but we have verified this symmetry numerically. Furthermore, the Fourier transform with respect to \(\omega\) can be inverted, and the expression of the four-point correlation function \(C_4(\kk_1, \kk_2, \kk_3, \kk_4; t_1, t_2, t_3, t_4)\), with times ordered as \(0 < t_1 < t_2 < t_3 < t_4\), is provided in the associated Python notebook \cite{dataset}.
}

\section{Numerical simulations}
\label{supp_numerical}

\begin{figure}
	\begin{center}
	\includegraphics[width=0.4\columnwidth]{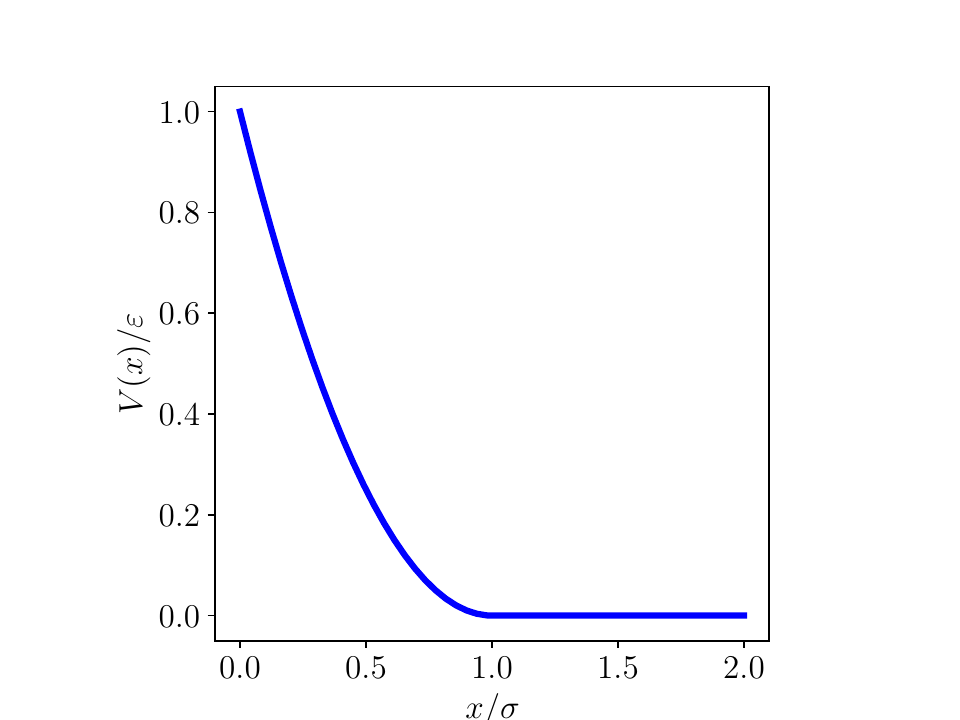}
	\caption{Plot of the interaction potential used in the Brownian dynamics simulations: $V(x) = \varepsilon (1-x/\sigma)^2\Theta(\sigma-x)$.}
	\label{fig_potential}
	\end{center}
\end{figure}

We performed Brownian dynamics simulations, i.e. time-integration of the overdamped Langevin equation given as Eq. \eqref{overdampedLangevin}. For a one dimensional system, the integrated equation reads:
\begin{equation}
	x_\alpha(t+\delta t) = x_\alpha(t) -\mu \, \delta t \sum_{\beta \neq \alpha} V'(|x_\alpha-x_\beta|) + \sqrt{2D\delta t} Z_{0,1}
	\label{eqBD}
\end{equation}
where $Z_{0,1}$ is a random number drawn from a normal distribution of mean 0 and variance 1. The $N=100$ particles are initially placed uniformly on a line of length $L$ (which is chosen to attain the chosen density $\rho_0=N/L$). The particle interact via a harmonic repulsive potential, which reads: $V(x) = \varepsilon (1-x/\sigma)^2\Theta(\sigma-x)$, where $\Theta$ is the Heaviside function {(see Fig. \ref{fig_potential} for a plot of the interaction potential)}. Its derivative, which is needed to compute the forces in Eq. \eqref{eqBD}, reads $V'(x) = -2\varepsilon(1-x/\sigma)\Theta(\sigma-x)$. The length $\sigma$ is taken equal to $1$ and sets the unit length of the simulation. We also take $D=\mu=1$ (this sets the typical diffusive timescale $\sigma^2/D$ to 1), and $\varepsilon=0.5 \kB T$. The integration timestep is $\delta t=0.01$.

The simulation is initally run for $10^4$ steps, for equilibration. The correlation functions shown in the main text are computed on the next $10^4$ steps: at each time step, the Fourier transform of the microscopic density is computed as $\tilde \rho(k,t) = \sum_{\alpha=1}^N \ex{-\ii k x_\alpha(t)}$. Importantly, since the system has periodic boundary conditions, the wavevectors $k$ must be chosen as multiples of $2\pi/L$ \cite{Kneller1995}. The structure factors and correlation functions are computed directly from their definition in the main text in terms of the functions $ \tilde \rho(k,t) $, and averages are performed over initial conditions and noise realisations (typically $2\cdot 10^4$ independent realizations).

{ Finally, we emphasize that analytical calculations are performed in the thermodynamic limit where $N$, $L\to\infty$ with a fixed density $\rho_0=N/L$. Results for finite-size systems, which are required to make comparisons with numerical simulations, can be obtained by making the change $\delta(k)\to(\mathcal{V}/(2\pi)) \delta_{k,0}$.}

\twocolumngrid

%\bibliography{biblio_nonGaussianDK}


\end{document}